\documentclass{PoS}

\title{A Model of High Energy Activity of Millisecond Pulsars in Globular Clusters}

\ShortTitle{A Model of High Energy Activity of Millisecond Pulsars in Globular Clusters}

\author{\speaker{Anna Zajczyk}$^{ab}$, Bronek Rudak$^a$, Jarek Dyks$^a$\\
	\\
        \llap{$^a$}Nicolaus Copernicus Astronomical Center,\\ ul. Rabianska 8, 87-100 Torun, Poland\\
	\\
	\llap{$^b$}LPTA, Universite Montpellier II,\\ Place Eugene Bataillon, 34095 Montpellier Cedex 05, France\\
	\\
        E-mail: \email{zajczyk@ncac.torun.pl},  \email{bronek@ncac.torun.pl}, \email{jinx@ncac.torun.pl}}

\abstract{
Resluts of 3D numerical simulations in the framework of pair starved polar cap model (PSPC) for millisecond pulsars are presented. In the investigated PSPC model electric field structure highly depends on the pulsar inclination which is clearly visible in the computed spectra of electrons escaping pulsar magnetosphere which become bimodal with the increasing inclination. This is an important result for modelling very high energy radiation from globular clusters where currently the standard power law or monoenergetic electron distributions are used.
}

\FullConference{High Time Resolution Astrophysics (HTRA) IV - The Era of Extremely Large Telescopes\\
		 May 5 - 7, 2010 \\
		 Agios Nikolaos, Crete Greece}

\begin{document}
%------------------------------
\section{Introduction}
\label{intro}

Recent observations by the Fermi LAT have firmly established millisecond pulsars (MSPs) as a class of gamma-ray emitters. This long-awaited result is, however, accompanied by an astonishing finding: the radiation characteristics of all nine gamma-ray MSPs in terms of the phase-averaged spectra and the lightcurves are similar to the gamma-ray properties observed for classical pulsars. In particular, contrary to prior expectations, the results suggest that particle accelerators in MSPs are spatially extended in a way predicted by outer-gap and/or two-pole-caustic slot gap models. It is worth noting that currently all but one of the gamma-ray detected MSPs are pulsars from the field of the Galaxy. The exception is MSP B1821-24, belonging to M28 globular cluster, which was detected by AGILE space observatory in 2009 (\cite{agile09}).

On the other hand, a multitude of MSPs reside in the dense cores of globular clusters (GCs). Due to a different evolutionary history in comparison with the field MSPs, one may expect that MSPs in GCs can have different structure of the accelerating gap than the field ones. For this reason we investigate, following \cite{mh04} and \cite{venterB}, a millisecond pulsar radiation model where accelerating electric field extends from the polar cap surface up to very high altitudes in the volume filled with open magnetic field lines. The large spatial extent of the electric field is justified by a low number of created pairs, which is insufficient to effectively screen the electric potential. Such scenario is called a pair starved polar cap model (PSPC).

On the basis of recent Fermi results (\cite{abdo10}, \cite{kong10}), it has been suggested that cumulative high energy emission of MSPs found in globular clusters contributes to their overall gamma-ray radiation. In addition to photons, MSPs inject relativistic electrons into GCs environment. These electrons propagate in the cluster and interact with its magnetic field producing X-ray radiation. Moreover, they are expected to up-scatter ambient photon fields (cosmic microwave background, infrared background and starlight) to high (HE) and very high energies (VHE). Recent theoretical studies show that such ICS photons may substantially contribute to HE and VHE radiation from globular clusters (\cite{bednarek07}, \cite{venterA}, \cite{cheng10}).

In our study we investigate both the characteristics of HE and VHE radiation from MSPs and the distribution of relativistic electrons that escape pulsar magnetosphere. The results are based on a small sample of modelled MSPs, nonetheless, they show interesting characteristics of the ejected electron spectra which should be taken into account in modelling of the ICS component in GCs.

%------------------------------
\section{Numerical simulations}
\label{sim}

Calculations were performed using 3D numerical model of pulsar magnetosphere described in the framework of space charge limited flow. The magnetic field is in the form of a retarded vacuum dipole (\cite{dyks04}) for which the curvature radii of magnetic field lines are determined in the inertial frame of reference. Moreover, following \cite{dyks02} the special relativity effects like aberration and time-of-flight delays are treated accordingly throughout the calculations.

An important assumption is that acceleration of particles takes place in the whole volume determined by the open magnetic field lines (PSPC model, e.g. \cite{mh04}). The electrons are injected at the polar cap surface at the Goldreich-Julian rate with low initial energy.
The accelerating electric field is in the form proposed by \cite{mh04} which takes into account the effect of dragging of inertial frames. In order to describe the electric field that extends up to the light cylinder three different formulae determining $E_{||}$ at different heights above the pulsar surface are combined. At moderate heights and close to the light cylinder the electric field is described by
\begin{equation}
E_{\mathrm{far}} \simeq E_{2} \exp\left[-\frac{\eta-1}{\eta_{\mathrm{c}}-1} \right] + E_{3}
\label{eqtot}
\end{equation}
where $E_{2}$ is given by \textit{equation} (14) and $E_{3}$ by \textit{equation} (35) of \cite{hm98}, $\eta = r/R_{\mathrm{NS}}$, $r$ is a radial distance from the neutron star and $R_{\mathrm{NS}}$ is a neutron star radius. In this approach $\eta_{\mathrm{c}}$ is a radial parameter which is determined for each magnetic field line via matching procedure. Its value depends on pulsar parameters like a magnetic field strength at the polar cap $B_{\mathrm{pc}}$ and an inclination angle $\alpha$ (i.e. the angle between the magnetic axis and the rotation axis) and also on a position of the magnetic field line's foot-point at the polar cap. In the neutron star vicinity $E_{\mathrm{near}}$, given by \textit{equation} (12) of \cite{dyks00}, describes the electric field. We require the electric field $E_{||}$ to be negative for all $r$ up to the light cylinder. If this condition is not fulfilled for a certain magnetic field line, the line is left out from the calculations so no acceleration of particles takes place along this line.

Our treatment of the PSPC model is different from that of \cite{venterB} in the way the near electric field $E_{\mathrm{near}}$ is matched with the far electric field $E_{\mathrm{far}}$. \cite{venterB} first transit from $E_{\mathrm{near}}$ to $E_{2}$ at a specific $\eta$ equal to $\eta_{\mathrm{b}} \approx 1 + P^{-0.333}$. Then $E_{2}$ is matched with $E_{3}$ so for distances close to the light cylinder the electric field converges to $E_{3}$. 
To determine the electric field across the pulsar magnetoshpere we do not use the formula for $\eta_{\mathrm{b}}$. In our calculations the matching of $E_{\mathrm{near}}$ with $E_{\mathrm{far}}$ takes place where condition $E_{\mathrm{near}} \simeq E_{\mathrm{far}}$ is fulfilled. This results in a different electric field structure across the pulsar magnetosphere with respect to \cite{venterB}.

%------------------------------ 
\section{Results}
\label{res}

\begin{figure}
  \begin{tabular}{ll}
	\includegraphics[scale=0.4]{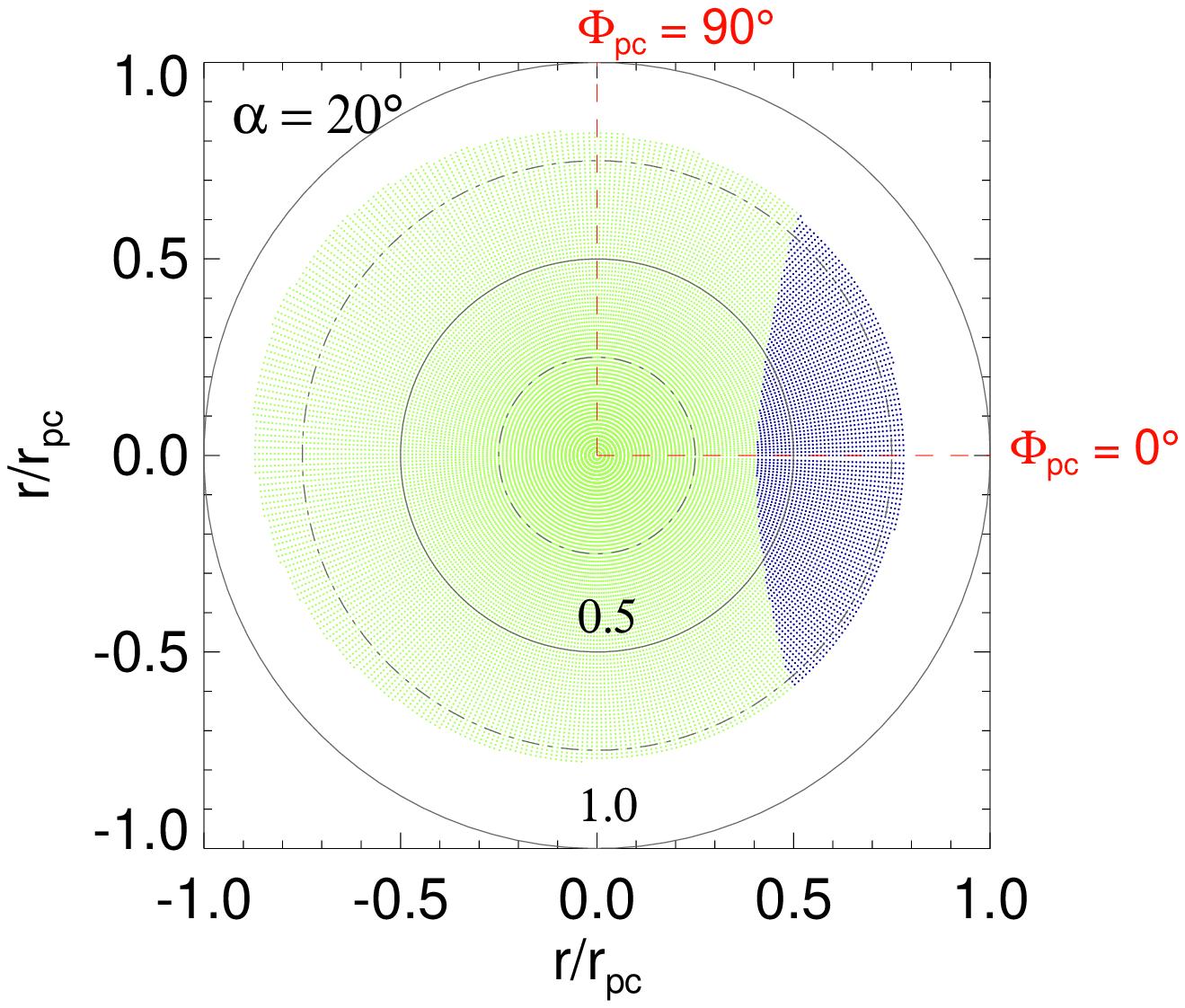} & \hspace{-0.5cm}\includegraphics[scale=0.4]{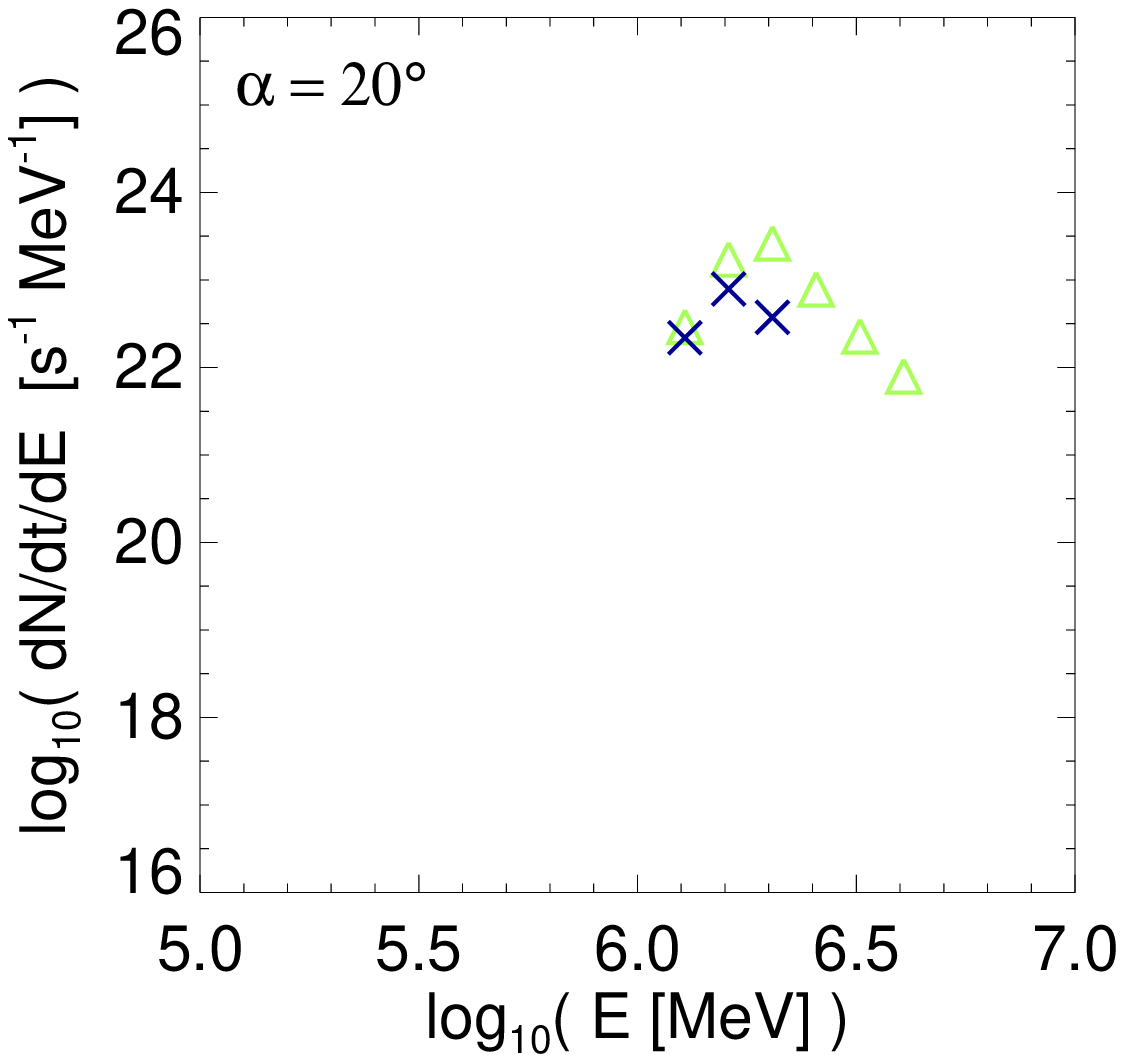}\\
	\includegraphics[scale=0.4]{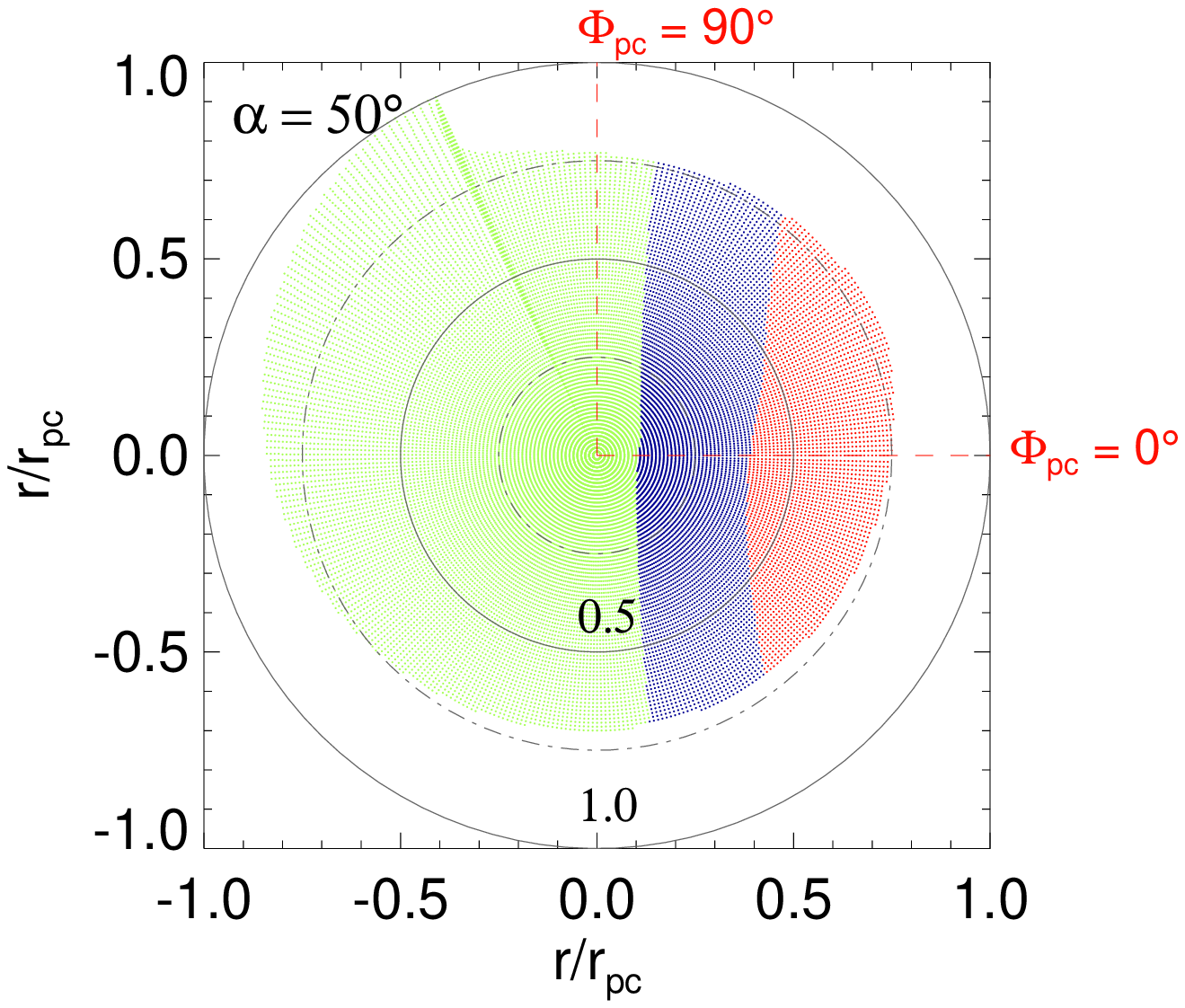} & \hspace{-0.5cm}\includegraphics[scale=0.4]{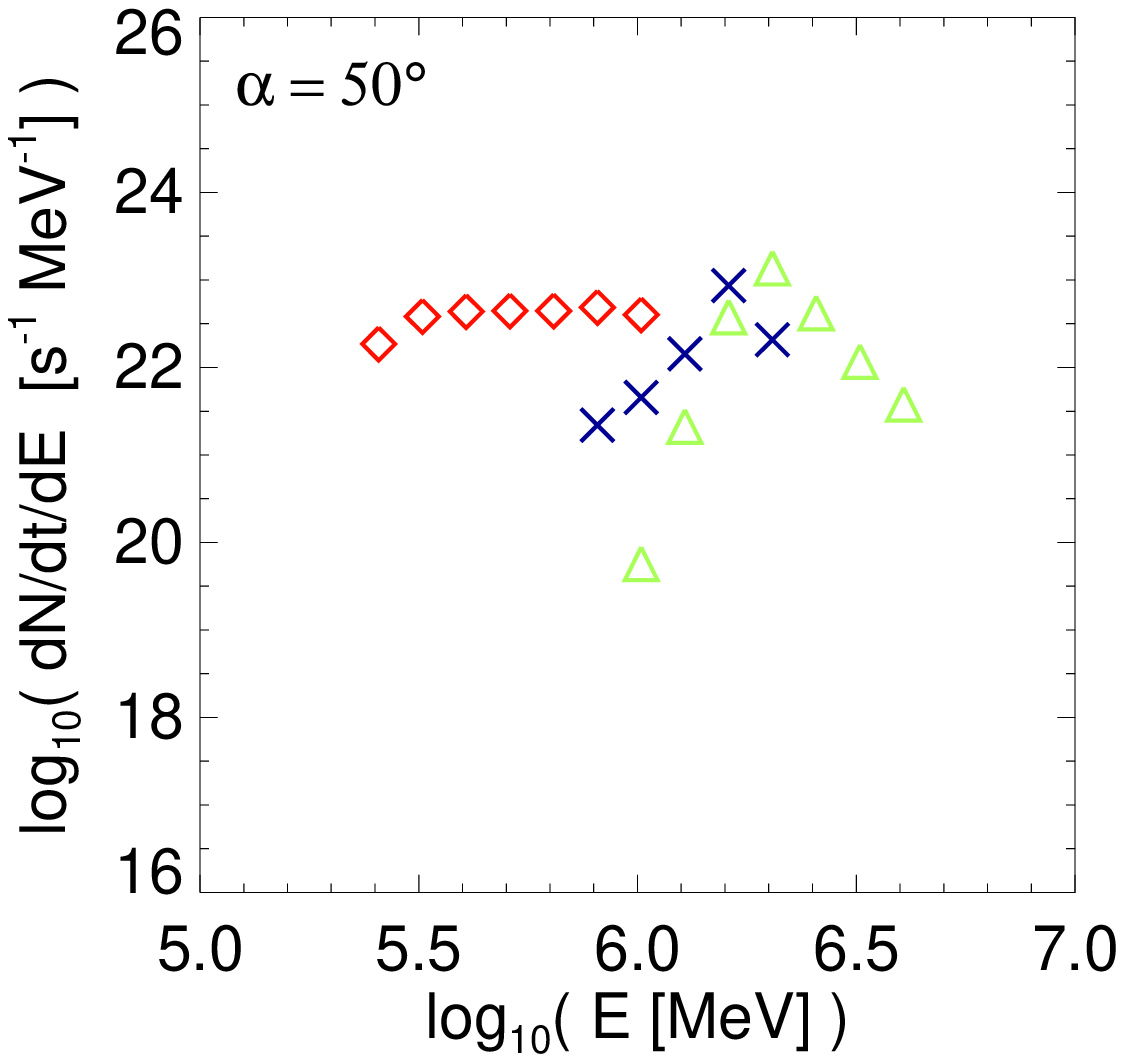}\\
	\includegraphics[scale=0.4]{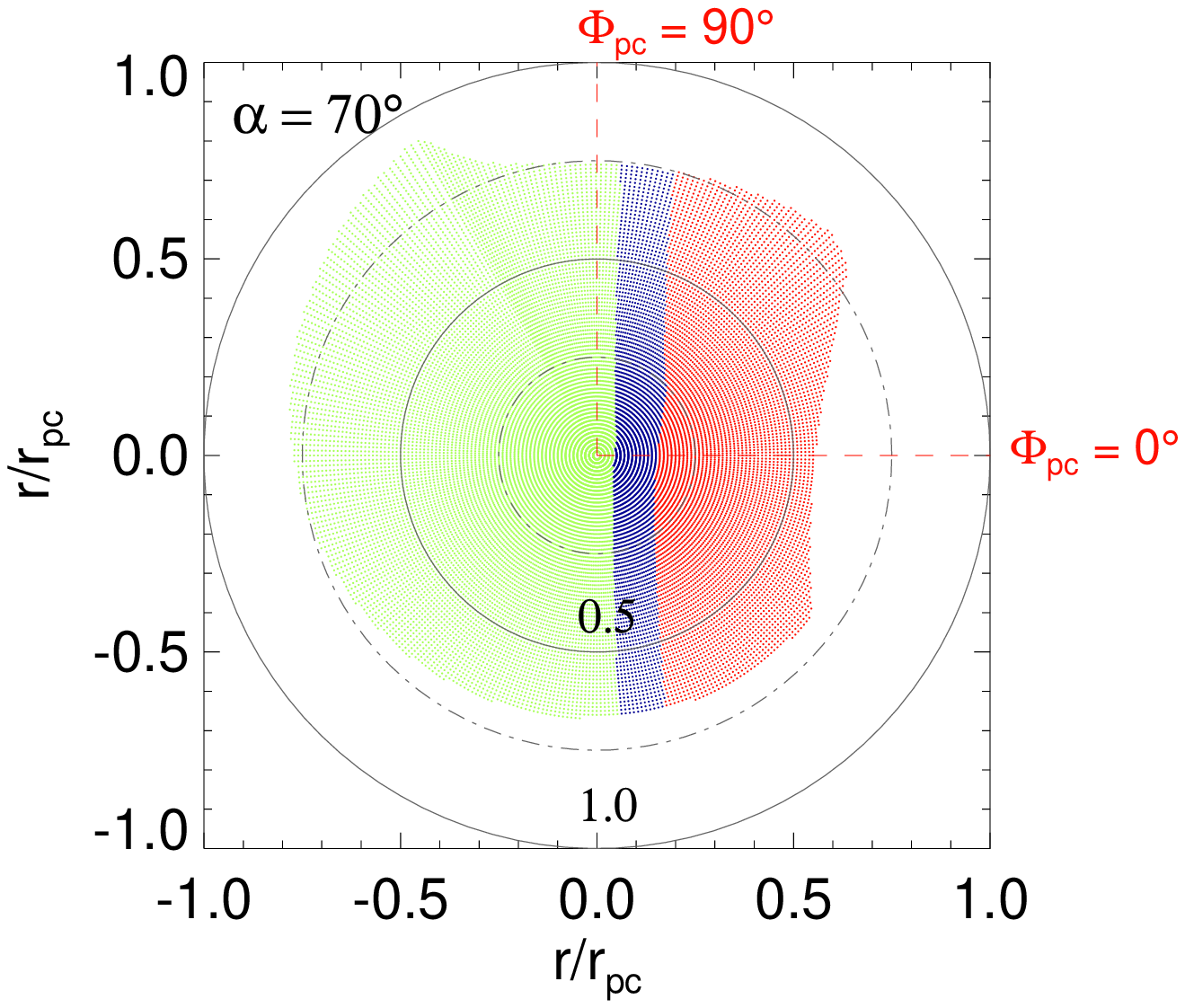} & \hspace{-0.5cm}\includegraphics[scale=0.4]{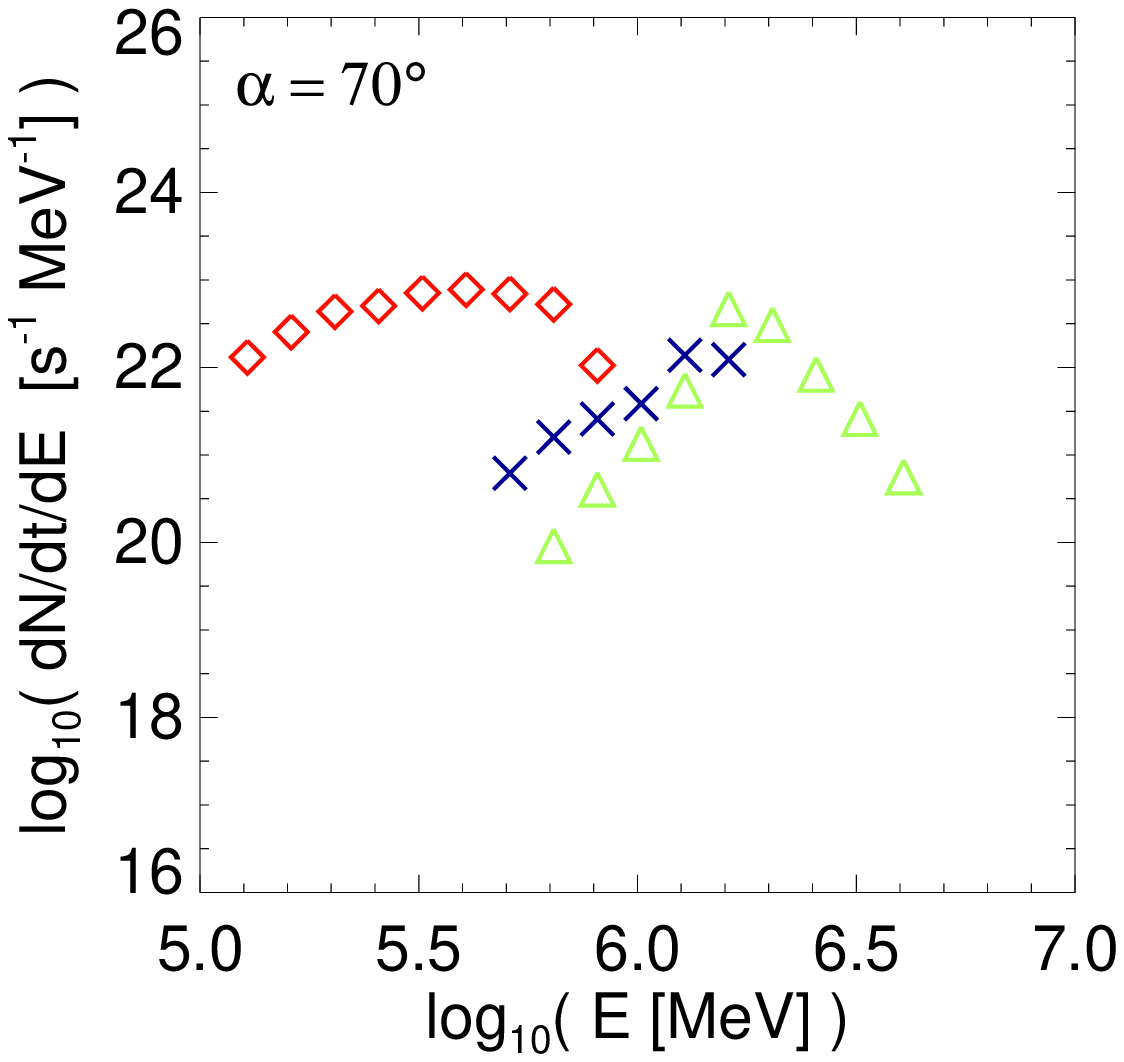}\\
  \end{tabular}
\caption{Presented plots are for MSP with $P=5$ ms, $B_{\mathrm{pc}}=3.5\times10^{8}$ G and the inclination angle $\alpha$, going from top to bottom: $20^{\circ}$, $50^{\circ}$ and $70^{\circ}$. \textit{Left panel:} schematic view of electron distribution across pulsar polar cap. Colour coded are different values of a parameter $\eta_{\mathrm{c}}$ used for combining the near electric field solution with the far solution (green = 5.6, red = 1.1, blue = all values between 1.1 and 5.6). Small and big solid circle mark the distance across the polar cap equal to $0.5r_{\mathrm{pc}}$ and $1.0r_{\mathrm{pc}}$, respectively. Here $r_{\mathrm{pc}}$ is the polar cap radius. Horizontal red dashed line shows $\Phi_{\mathrm{pc}}=0^{\circ}$ while vertical one points to $\Phi_{\mathrm{pc}}=90^{\circ}$ ($\Phi_{\mathrm{pc}}$ is the magnetic azimuth angle). \textit{Right panel:} spectra of electrons escaping MSP magnetosphere. Colour coding is the same as for the left panel and shows which electrons (i.e. propagating along which open magnetic field lines) contribute to different parts of the spectrum.}
\label{fig1}
\end{figure}

The values of the key parameter ($\eta_{\mathrm{c}}$) determining the shape of the accelerating electric field along open magnetic field lines are presented in the left panel of Figure~\ref{fig1} as a function of the line's foot-point at the polar cap. It is clearly visible that with the increasing pulsar inclination angle the structure of the electric field changes drastically. For small inclination angles most of the volume determined by the open magnetic field lines is filled with the electric field solutions described by high values of $\eta_{\mathrm{c}}$. As the pulsar inclination becomes higher the electric field solutions determined using low values of $\eta_{\mathrm{c}}$ start to appear on the part of the polar cap that is further away from the pulsar rotation axis (\textit{left panel} of Fig.~\ref{fig1}). For large $\alpha$ (e.g. $\alpha = 70^{\circ}$) there is a part of the open magnetosphere where electrons are not accelerated because no $\eta_{\mathrm{c}}$ is found to match the electric field formulae.

Such behaviour of the electric field structure with inclination angle has an influence on the spectrum of electrons escaping the millisecond pulsar magnetosphere. 
In the \textit{right panel} of Figure~\ref{fig1} different electron spectra are presented. For MSP with low inclination the electron spectrum is single peaked and the energies of electrons contributing to the peak are in the range from $10^{6}$ to $10^{6.7}$ MeV. The electron spectra of highly inclined MSPs are bimodal. As can be seen from Figure~\ref{fig1} electrons that are accelerated along lines with high value of the parameter $\eta_{\mathrm{c}}$ contribute to a high energy component (the same as for the low inclination MSPs) while those whre $\eta_{\mathrm{c}}$ is low build up a low energy component that streaches down to energies of $10^{5}$ MeV.

\begin{figure}
  \begin{tabular}{lll}
	\includegraphics[scale=0.3]{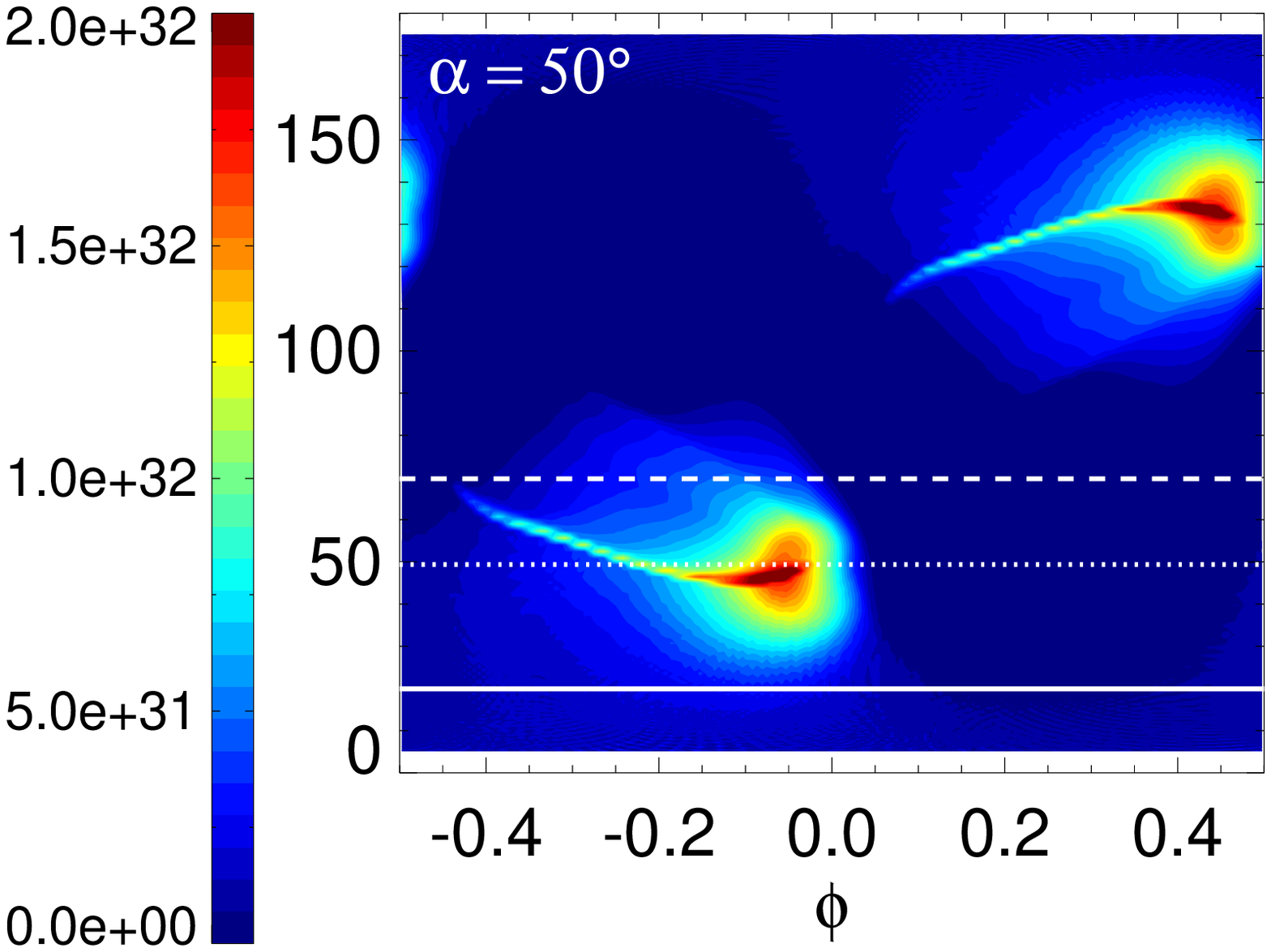} & \hspace{-0.5cm}\includegraphics[scale=0.3]{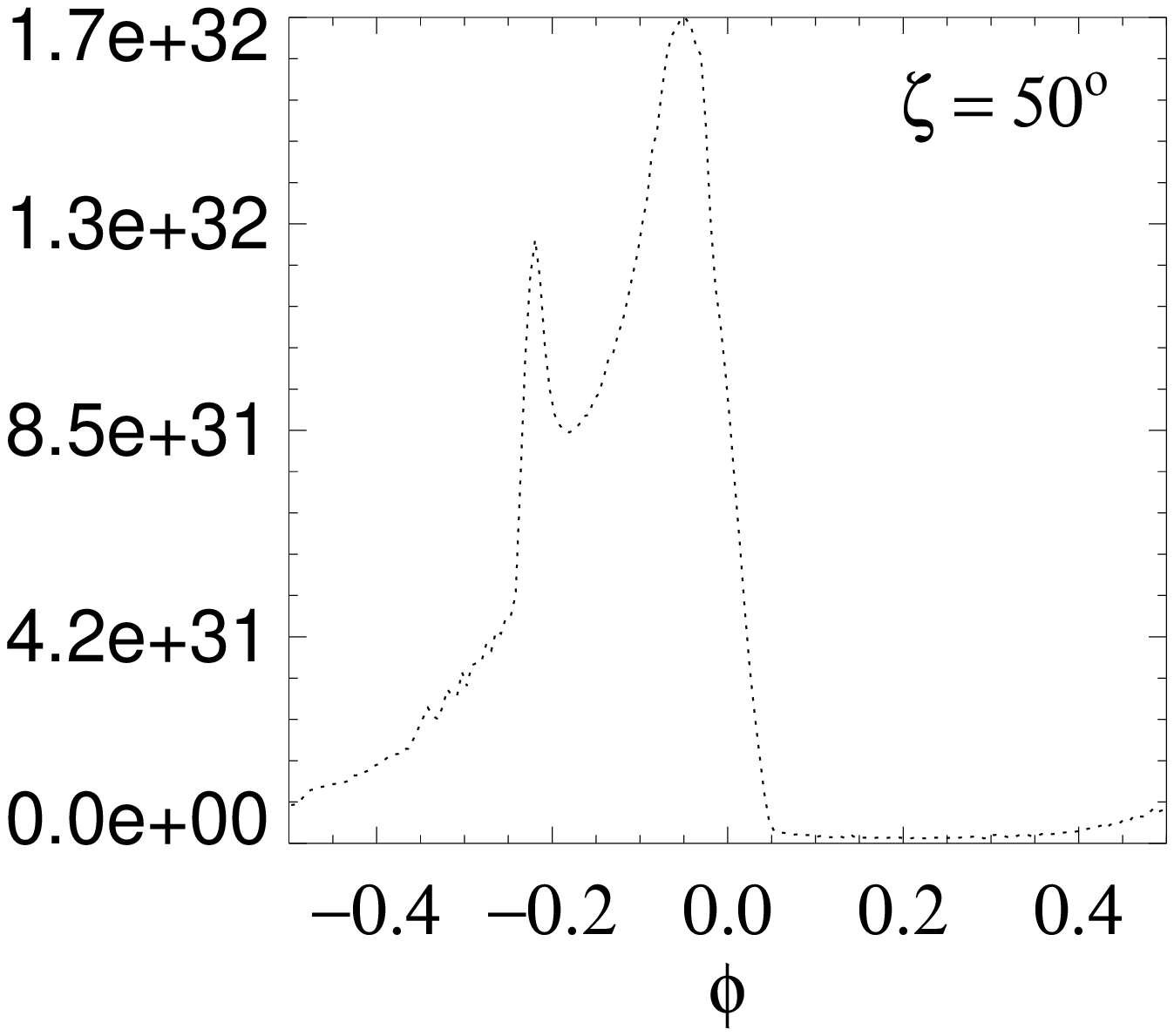}&\hspace{-1.0cm}\includegraphics[scale=0.3]{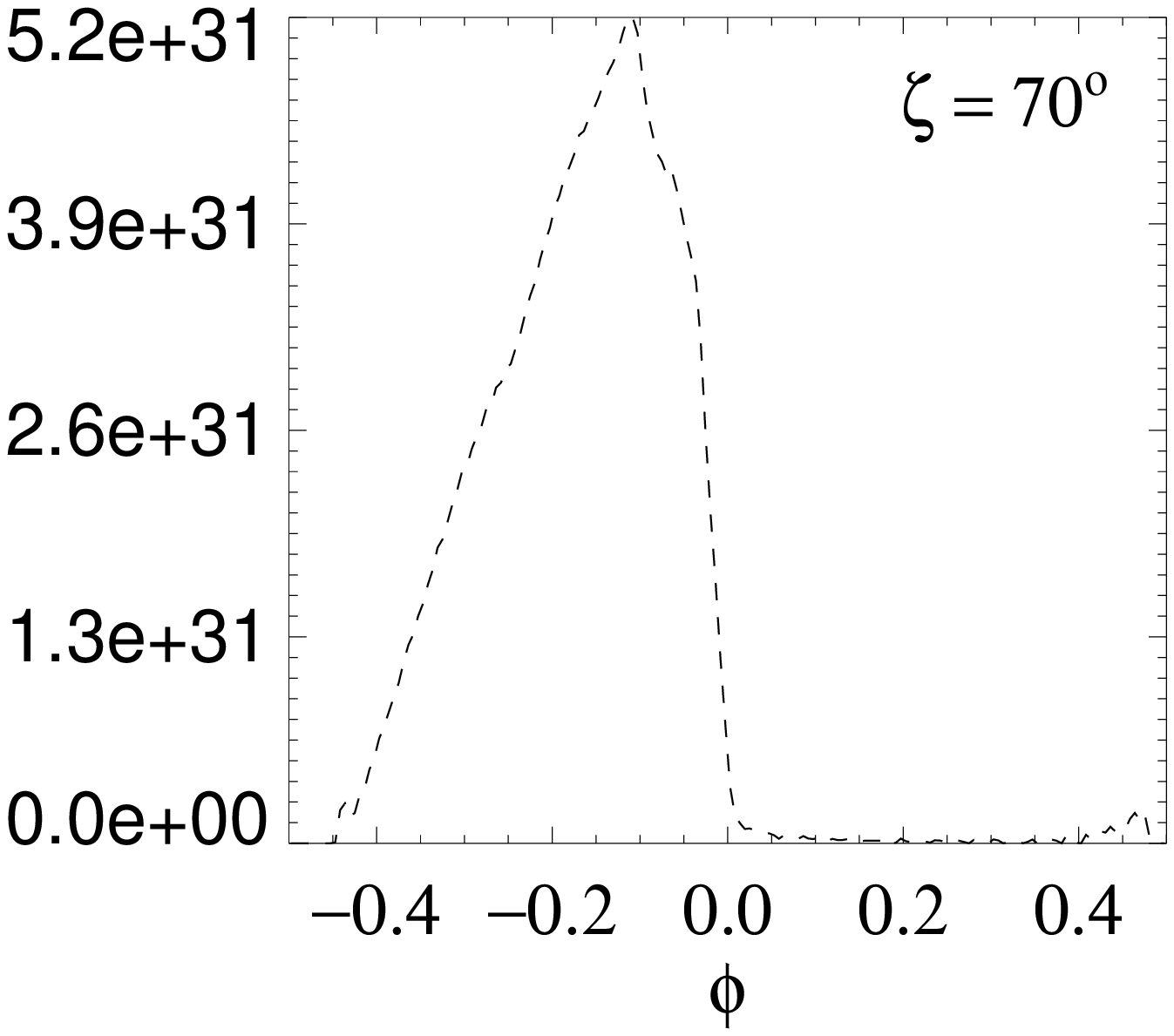}\\
	\includegraphics[scale=0.3]{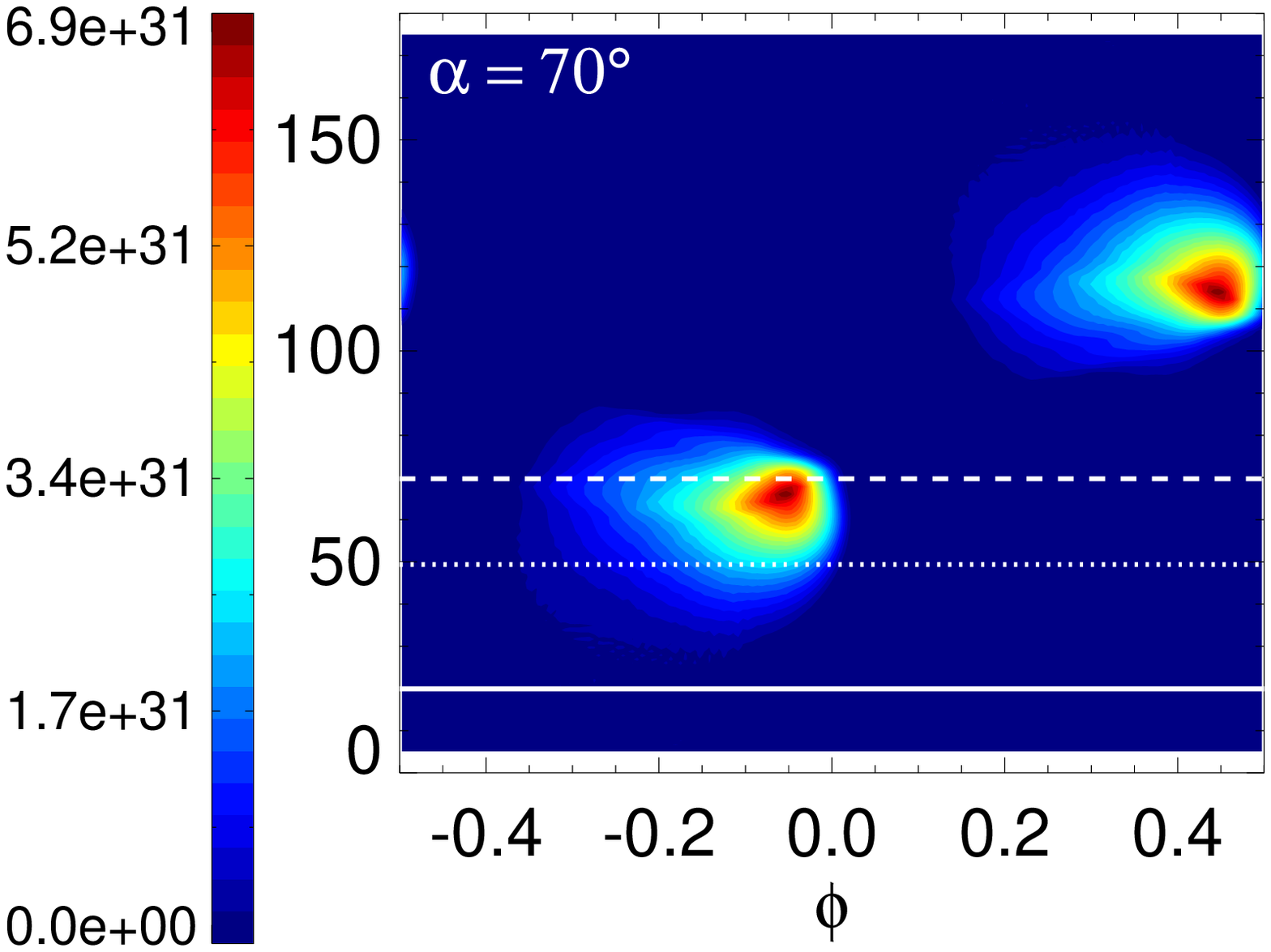} & \hspace{-0.5cm}\includegraphics[scale=0.3]{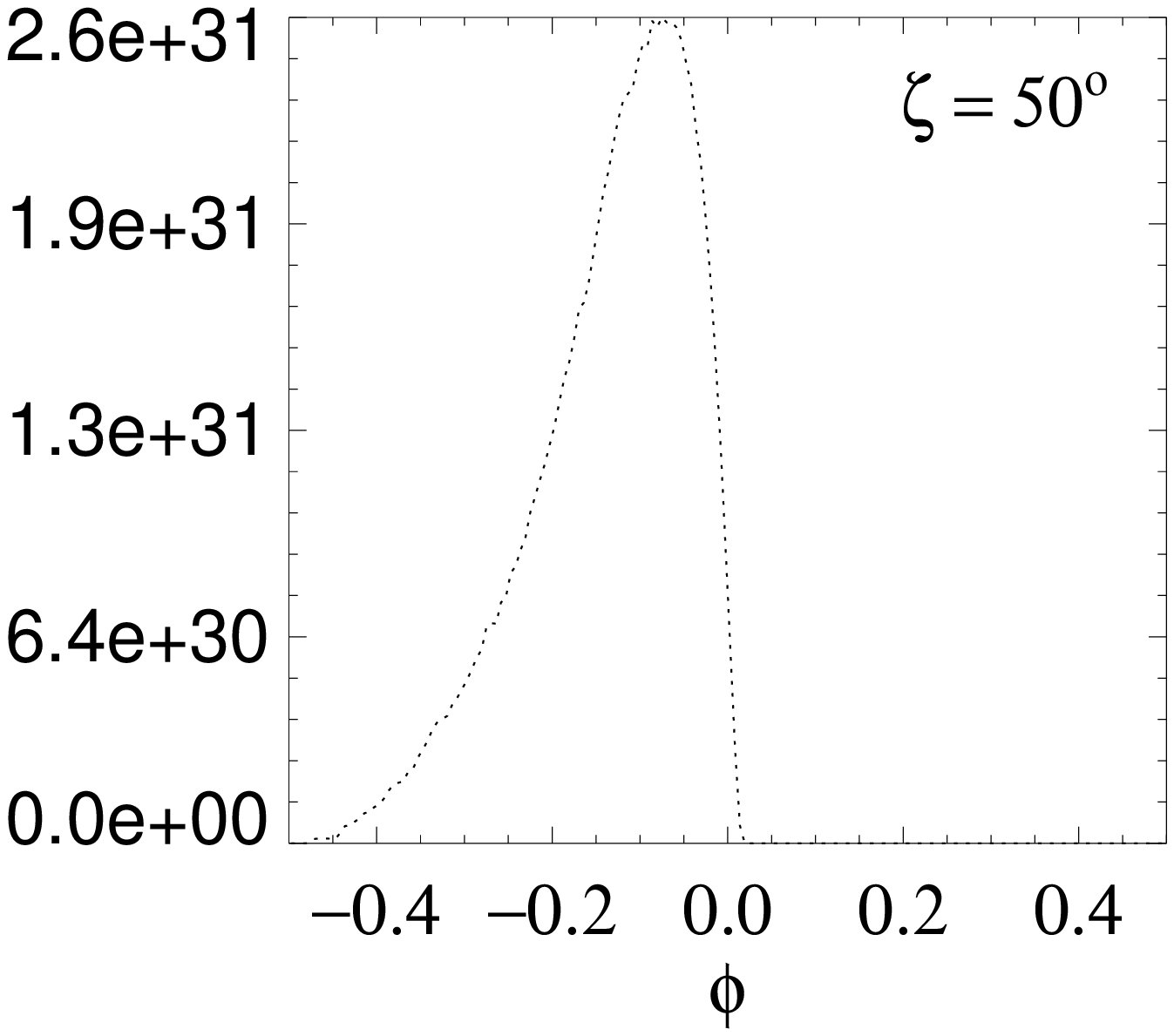}&\hspace{-1.0cm}\includegraphics[scale=0.3]{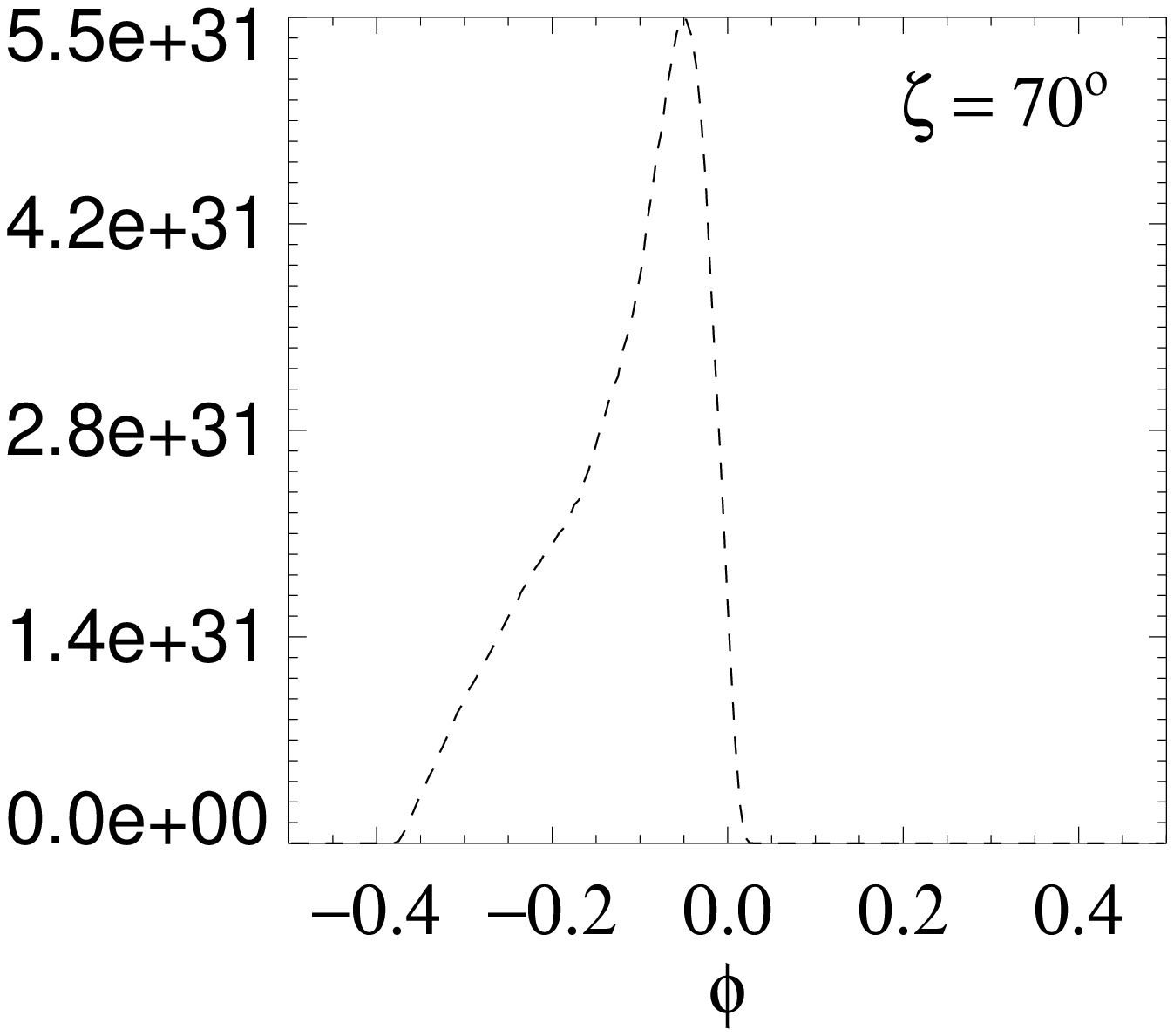}\\
  \end{tabular}
\caption{Photon maps and light curves obtained for MSP with $P=5$ ms and $B_{\mathrm{pc}}=3.5\times10^{8}$ G. Top panel shows pulsar with inclination of $50^{\circ}$, while bottom panel shows the case for $70^{\circ}$. In the photon maps a photon density is colour coded in arbitrary units. On the x-axis a pulsar rotation phase $\phi$ is presented while the y-axis shows a viewing angle $\zeta$ in degrees. Pulse profiles for observers with viewing angles $\zeta = 50^{\circ}$ and $70^{\circ}$ are presented.}
\label{fig2}
\end{figure}

The electric field structure affects also the photon characteristics of MSPs. Figure~\ref{fig2} shows photon maps and the light curves for selected observers in the energy range between 100MeV and 30GeV. Within the framework of PSPC model one obtains single pulse lightcurves. In the photon map for a pulsar with the inclination of $50^{\circ}$ a wing-like feature is present. This corresponds to emission from the magnetic field lines anchored within the polar cap notch.

%------------------------------ 
\section{Conclusions}
\label{concl}

Our studies give a new insight into possible spectra of relativistic electrons that are ejected from the millisecond pulsar magnetospheres into globular cluster. The spectra are bimodal provided $\alpha > 20^{\circ}$, which is a new result with respect to the standard power law or monoenergetic electron distributions that are currently used in modelling of the VHE emission of globular clusters. 

%------------------------------
\acknowledgments
AZ wishes to thank the organizers of HTRA4 for financial support. 
This research is supported by the MNiSW grant N203387737.

%------------------------------

\end{document}